\documentclass[twocolumn,prb,aps,nobalancelastpage,amssymb,showpacs]{revtex4}

\usepackage{graphicx}

\begin{document}
\title{Fermi-surface reconstruction and two-carrier model for the Hall effect in YBa$_2$Cu$_4$O$_8$ }

\author{P.~M.~C.\ Rourke,$^1$ A.~F.\ Bangura,$^1$ C.\ Proust,$^{2,3}$ J.\ Levallois,$^2$ N.\ Doiron-Leyraud,$^4$ D.\ LeBoeuf,$^4$ L.\ Taillefer,$^{3,4}$ S.\ Adachi,$^5$ M.~L.\ Sutherland,$^6$ and N.~E.\ Hussey$^1$}

\affiliation{$^1$H. H. Wills Physics Laboratory, University of Bristol, Tyndall Avenue, Bristol, BS8 1TL, United Kingdom}
\affiliation{$^2$Laboratoire National des Champs Magn\'{e}tiques Intenses, CNRS, Toulouse, France}
\affiliation{$^3$Canadian Institute for Advanced Research, Toronto, Ontario, Canada M5G 1Z8}
\affiliation{$^4$D\'{e}partement de Physique and RQMP, Universit\'{e} de Sherbrooke, Sherbrooke, Quebec, Canada J1K 2R1}
\affiliation{$^5$Superconductivity Research Laboratory, International Superconductivity Technology Center, Shinonome 1-10-13, Koto-ku, Tokyo 135-0062, Japan}
\affiliation{$^6$Cavendish Laboratory, University of Cambridge, Madingley Road, Cambridge CB3 0HE, United Kingdom}

\date{\today}

\begin{abstract}
Pulsed field measurements of the Hall resistivity and magnetoresistance of underdoped YBa$_2$Cu$_4$O$_8$ are analyzed self-consistently using a simple model based on coexisting electron and hole carriers. The resultant mobilities and Hall numbers are found to vary markedly with temperature. The conductivity of the hole carriers drops by one order of magnitude below 30 K, explaining the absence of quantum oscillations from these particular pockets. Meanwhile the Hall coefficient of the electron carriers becomes strongly negative below 50 K. The overall quality of the fits not only provides strong evidence for Fermi-surface reconstruction in Y-based cuprates, it also strongly constrains the type of reconstruction that might be occurring.
\end{abstract}

\maketitle

A key step to unraveling the mystery of high-temperature superconductivity in the cuprates is an elucidation of the nature of the normal state from which the superconductivity emerges and its evolution with doping. At half filling, the cuprate plane is insulating and antiferromagnetically ordered. With increasing (hole) doping, antiferromagnetism is suppressed and superconductivity first rises, then diminishes. Beyond the superconducting dome, a highly correlated Fermi liquid with a large Fermi surface (FS) is recovered.\cite{Hussey08} The nature of the intermediate state lying between these two extremes however remains an outstanding and controversial issue.

Angle-resolved photoemission spectroscopy (ARPES) suggests the existence of Fermi arcs, disjointed lines of low-lying excitations centered along the zone diagonals, in the underdoped (UD) regime.\cite{Norman98, Shen05, Hussain08} The nodal arc picture implies a gradual destruction of the FS (or decoherence of quasiparticle states) with increasing correlation strength due to the formation of an anisotropic pseudogap that is peaked at the zone boundaries.

Quantum oscillation (QO) experiments, meanwhile, suggest a very different scenario for the ground state of UD cuprates involving closed {\it pockets}, not lines, of coherent excitations.\cite{Doiron-Leyraud07, Yelland08, Bangura08, Jaudet08, Sebastian08, Audouard09, Sebastian09} Coupled with the observation of a negative Hall effect $R_{\rm H}$ (Ref.~\onlinecite{LeBoeuf07}) and thermopower\cite{Chang10} at low $T$, a picture then emerges in which the large holelike FS seen on the overdoped (OD) side undergoes symmetry-breaking reconstruction into electron ($e$) and hole ($h$) pockets or sheets at a critical doping level $p_c$.\cite{Taillefer09}

A number of competing states have been proposed as the origin of FS reconstruction in UD cuprates, including stripes,\cite{MillisNorman07} commensurate,\cite{Sachdev09} and incommensurate\cite{Harrison09} spin-density waves (SDWs) and $d$-density wave ($d$DW) order.\cite{ChakravartyKee08} Such scenarios are conceptually incompatible with the notion of Fermi arcs and while alternative theoretical\cite{Harrison07, Pereg-Barnea09} and experimental\cite{Meng09} pictures have emerged that seek to reconcile the conflicting outcome of ARPES and QO studies, several key questions remain, not least concerning the origin of the QO themselves.

One universal feature of models based on FS reconstruction is the prediction of multiple pockets and corresponding multiple frequencies in the QO spectra.  However, in most measurements carried out to date, only one frequency is routinely observed\cite{Doiron-Leyraud07, Bangura08, Yelland08, Jaudet08, Helm09} (other than those due to splitting caused by $c$-axis dispersion), despite significant improvements in signal to noise.\cite{Audouard09} Moreover, the observation of a negative $R_{\rm H}$ (Ref.~\onlinecite{LeBoeuf07}) implies that $e$-like carriers are the most mobile, even though the reconstruction scenarios invariably locate the $e$-pocket near the zone boundaries where, according to ARPES, the pseudogap is maximal and all states are incoherent.

In order to help resolve these controversies, we have carried out a detailed analysis of the low-$T$ magnetoresistance (MR) and Hall resistivity of UD YBa$_2$Cu$_4$O$_8$ (Y124) using the simplest workable model of a metal containing electron-like and hole-like carriers. Self-consistent field-independent fitting parameters are obtained at all temperatures and for all fields above a notional ($T$-dependent) cut-off field $H_n$ that corresponds to the restoration of the resistive normal state (a detailed discussion of $H_n(T)$ can be found in the supplementary information of Ref.~\onlinecite{LeBoeuf07}). This level of consistency provides strong evidence in favor of FS reconstruction in the Y-based cuprates Y124 and (by inference) YBa$_2$Cu$_3$O$_{7-\delta}$ (Y123). The data also reveal a remarkable evolution in both the carrier mobility and the Hall coefficient of the individual pockets that place strong constraints on the actual reconstruction that might be present.

Y124 single crystals ($T_c = 80$~K) were mounted for simultaneous longitudinal ($\rho_{xx}$) and Hall ($\rho_{xy}$) resistivity measurements, as described in Ref.~\onlinecite{Bangura08}, and placed in a pumped $^4$He cryostat at the LNCMI-T pulsed field facility, with the field $\mathbf{H}||c$. In Y124, the in-plane resistivity is anisotropic due to the (significant) contribution of the double CuO chains aligned along the $b$ axis. While $\rho_{xy}$ is found to be identical for both current directions, as expected from the Onsager relation, we consider here only data taken with $\mathbf{I}||a$, in which configuration the CuO$_2$ plane contribution dominates in $\rho_{xx}$.

\begin{figure}
\includegraphics[width=7.0cm,keepaspectratio=true]{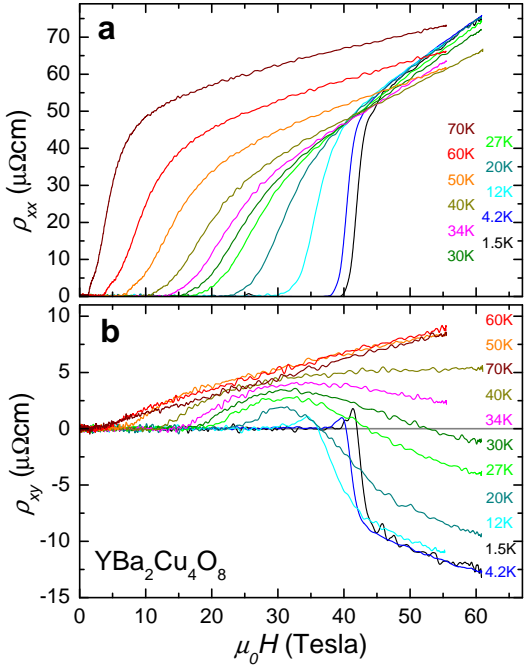}
\caption{(Color online) Magnetoresistance (top panel) and Hall resistivity (bottom panel) of YBa$_2$Cu$_4$O$_8$ measured at various temperatures (labeled) in pulsed fields up to 60 T.} \label{Fig1}
\end{figure}

Figure \ref{Fig1} shows a series of $\rho_{xx}$ (top panel) and $\rho_{xy}$ (bottom panel) curves taken on one Y124 single crystal (\#1) at selected temperatures between 1.5 and 70 K (similar data, summarized in Fig.~3, were recorded on a second crystal \#2). As reported previously,\cite{LeBoeuf07} $\rho_{xy}(H)$ is linear and positive at high $T$, nonlinear at intermediate $T$ and finally, linear and negative at the lowest temperatures. Similar behavior, with inverted sign, has also been seen in electron-doped Pr$_{2-x}$Ce$_x$CuO$_4$.\cite{Li07} The high-field MR, on the other hand, is approximately linear in $H$, in contrast with the quadratic MR response seen in OD La$_{2-x}$Sr$_x$CuO$_4$ (LSCO).\cite{Cooper09}

The high-field behavior of $\rho_{xx}$ and $\rho_{xy}$ in Y124 is characteristic of a metal containing two mobile carrier types. Accordingly, we proceed to model both components using a simple two-carrier Drude model with field-independent electron and hole carrier densities and mobilities,\cite{AshcroftMermin}
\begin{eqnarray}
\rho_{xx}(H) = \frac{(\sigma_h+\sigma_e) + \sigma_h\sigma_e(\sigma_hR_h^2+\sigma_eR_e^2)H^2}{(\sigma_h+\sigma_e)^2 + \sigma_h^2\sigma_e^2(R_h+R_e)^2H^2}, \label{eq:one}
\end{eqnarray}
\begin{eqnarray}
\rho_{xy}(H) = \frac{\sigma_h^2 R_h + \sigma_e^2 R_e + \sigma_h^2\sigma_e^2R_hR_e(R_h+R_e)H^2}{(\sigma_h+\sigma_e)^2 + \sigma_h^2\sigma_e^2(R_h+R_e)^2H^2}H, \label{eq:two}
\end{eqnarray}
where $\sigma_{e(h)}$ and $R_{e(h)}$ are the conductivity and Hall coefficient of the $e$($h$) carriers, respectively. Note the interdependence of $\rho_{xx}$ and $\rho_{xy}$ on the four fitting parameters and the equivalence of the denominator in both expressions. While both prove highly constraining when fitting $\rho_{xx}$ and $\rho_{xy}$ simultaneously, the error bars in $R_{e(h)}$ can be large when $\sigma_{e(h)}$ is small.

In order to extract $\rho_{xy}^{\rm pl}$, the {\it intrinsic} in-plane Hall resistivity in Y124, the shorting effect of the CuO chains on the Hall response also needs to be taken into account. This is achieved by multiplying the measured $\rho_{xy}$ by the anisotropy in the in-plane conductivities, i.e., $\rho_{xy}^{\rm pl} = \rho_{xy} \times (\rho_a/\rho_b)$,\cite{Segawa04} where $1/\rho_b = 1/\rho_a + 1/\rho_{\rm chain}$ and $\rho_{\rm chain} = \rho_0^{\rm ch} + 0.0015T^2$~$\mu\Omega$cm, as determined previously on the same batch of crystals.\cite{Hussey97} As we do not know precisely $\rho_0^{\rm ch}$, the residual resistivity on the CuO chains (since the samples are mounted for $\mathbf{I}||a$), we apply a least-squares fitting routine to both $\rho_{xx}(H)$ and $\rho_{xy}^{\rm pl}(H)$ (above $H_n$) for different values of $\rho_0^{\rm ch}$ until eventually we arrive at a full, self-consistent set of fitting parameters. Note that $\rho_0^{\rm ch}$ takes a single value for all the fits.

\begin{figure}
\includegraphics[width=8.0cm,keepaspectratio=true]{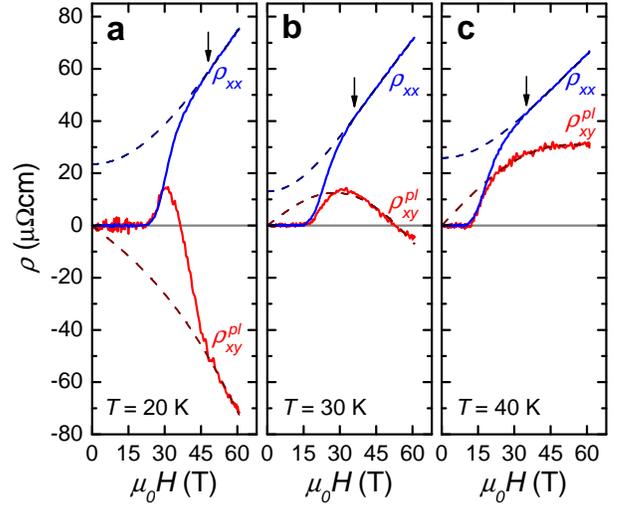}
\caption{(Color online) $\rho_{xx}(H)$ and $\rho_{xy}^{\rm pl}(H)$ data (solid) and fits (dashed lines) at (a) 20 K, (b) 30 K, and (c) 40 K. The arrows indicate $H_n$ [see Ref.~\onlinecite{LeBoeuf07} for a detailed discussion of $H_n(T)$].}
\label{Fig2}
\end{figure}

Figure 2 shows simultaneous fits of $\rho_{xx}(H)$ and $\rho_{xy}^{\rm pl}(H)$ in the temperature regime ($20 \leq T \leq 40$~K) where the Hall response undergoes the most marked changes with $T$ and $H$. In these fits, $\rho_0^{\rm ch} = 3$~$\mu\Omega$cm. Given the number of simplifying assumptions (field-independent parameters, no bilayer splitting, no intrinsic Hall and MR contributions from the CuO chains, and no Zeeman terms in the in-plane MR), the quality of the fits is remarkably good (the fits to the 27 and 30 K data are actually the worst in the whole data set, a point that we shall return to later). In our opinion, the highly constrained nature of the fits confirms that the normal state is being accessed above $H_n$ and constitutes some of the strongest evidence to date for FS reconstruction in UD cuprates.

Before discussing the resultant parameterization, we first show that our assumption to neglect the chain contribution to the MR is a reasonable one. To do this, we consider isostructural Pr124, in which only the CuO chains are metallic down to low $T$.\cite{Hussey02} First, for $T < 100$~K, $R_{\rm H} \leq 0.2$~mm$^3$/C,\cite{Horii02} two orders of magnitude smaller than in Y124. Second, assuming parallel plane ($\rho_a^{\rm pl}$) and chain ($\rho_a^{\rm ch}$) resistivities in Y124, the total $a$-axis MR is given by the weighted sum of the two individual contributions, i.e.,  $\Delta \rho_a/\rho_a = (\rho_a/\rho_a^{\rm ch})(\Delta \rho_a^{\rm ch}/\rho_a^{\rm ch}) + (\rho_a/\rho_a^{\rm pl})(\Delta \rho_a^{\rm pl}/\rho_a^{\rm pl})$.\cite{Hussey98} In Pr124 at $T = 4.2$~K, $\rho_a^{\rm ch}(\mathbf{I}||a) \sim 4$~m$\Omega$cm and $\Delta \rho_a^{\rm ch}/\rho_a^{\rm ch}(\mathbf{I}||a) \sim 4$ at 60 T (Ref.~\onlinecite{Narduzzo06}) while here, $\rho_a \sim 25$~$\mu\Omega$cm and $\Delta \rho_a/\rho_a \sim 3$. Since $\rho_a/\rho_a^{\rm ch} \sim 1/160$ and $\Delta \rho_a/\rho_a \sim \Delta \rho_a^{\rm ch}/\rho_a^{\rm ch}$, the chain contribution to $\Delta \rho_a/\rho_a$ can be safely neglected.

\begin{figure}
\includegraphics[width=7.0cm,keepaspectratio=true]{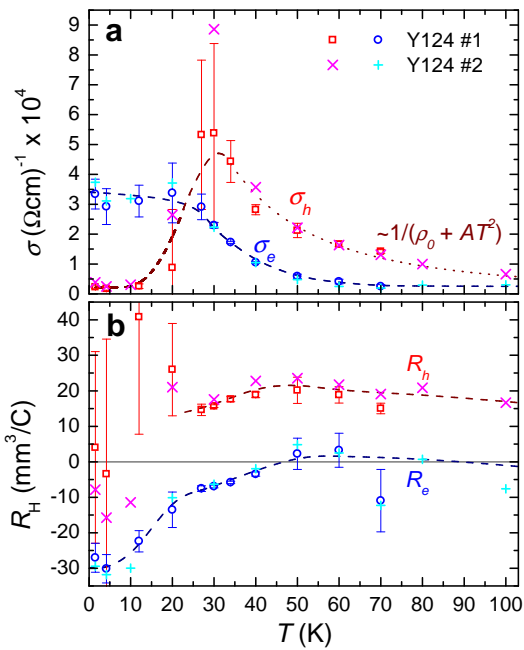}
\caption{(Color online) $T$-dependence of (a) $\sigma_e$ and $\sigma_h$ and (b) $R_e$ and $R_h$, the conductivities and Hall coefficients of the electron ($e$) and hole ($h$) pockets, derived from fits of $\rho_{xx}$ and $\rho_{xy}$ to Eqs. (1) and (2). The dotted line in (a) is a fit to 1/($\rho_0 + AT^2$). The dashed lines are guides to the eye.} \label{Fig3}
\end{figure}

Figure 3 shows the four fitting parameters, $\sigma_{e(h)}$ (top panel), and $R_{e(h)}$ (bottom panel), for the two $a$-axis crystals \#1 and \#2. The overall consistency of the fitting procedure implies that there is no {\it field-induced} FS reconstruction in Y124 or at least no further changes in the carrier properties above $H_n(T)$. The fact that a two-carrier model is sufficient to describe $\rho_{xx}$ and $\rho_{xy}$ in Y124 also implies that the Fermi surfaces of the bonding and antibonding bands in Y124 are rather similar and their interband coupling correspondingly weak. [Were this not the case, it would be essentially impossible to fit the data in any meaningful way with Eqs. (1) and (2)].

The most striking aspect of Fig.~3 is the dramatic collapse in $\sigma_h$ below a temperature $T_h \approx 30$~K. Above $T_h$, $\sigma_h > \sigma_e$, as one might anticipate from Luttinger's theorem for a hole-doped cuprate. In this region, $\sigma_h \sim 1/(A + BT^2)$ (see dashed line), as found for Y123 above 50 K.\cite{RullierAlbenque07} Below $T_h$, $\sigma_h$ drops (almost exponentially) by at least one order of magnitude, and at the lowest $T$, $\sigma_h$ is ten times smaller than $\sigma_e$. Because the error bars in $R_h$ are large below $T_h$, due to the small value of $\sigma_h$, we cannot tell how $R_h$ behaves below 20~K. Clearly though, $T_h$ marks the onset of a major change in either the Fermi-surface properties or the scattering rate that affects principally the mobility of the $h$-like carriers. As mentioned above, the fit to $\rho_{xy}(H)$ near $T = T_h$ is significantly worse than at any other temperature and may imply that the hole concentration and/or the mobility are changing with increasing field near the transition.

From the value of $\rho_0^{\rm ch}$, we obtain an intrachain mean-free-path of $\ell_{\rm ch} \sim 500$~\AA. Assuming an equivalent $\ell_{\rm pl}$ for the plane carriers and taking our value of $\sigma_e$ at low $T$, we obtain an estimate for the Fermi wave vector for 1 (2 identical) $e$ pocket(s) of $k_F^e = 2.2$(1.1)~nm$^{-1}$. From $R_e$(1.5~K) $= -30$~mm$^3$/C, we get a similar estimate of 1.4(1.0)~nm$^{-1}$. These values compare well with the value of 1.4~nm$^{-1}$ determined directly from QO experiments,\cite{Yelland08, Bangura08} confirming that the QO do indeed originate from cyclotron motion around the $e$ pocket and thus there is no need to invoke any other, more exotic explanation. Note too that the value of $R_e$ at $T \rightarrow 0$, having corrected for the chain contribution, is similar to that found directly in UD Y123.\cite{LeBoeuf07} The closeness in frequency of the QO from the two $e$ pockets in Y123 has been demonstrated recently through the observation of beating,\cite{Audouard09} while the absence in our experiments of any QO from the holes\cite{Doiron-Leyraud07, Bangura08, Jaudet08, Audouard09} is now understandable, at least empirically, in terms of the marked decrease in the hole mobility below $T_h$.

At $T \sim 50$~K and above, $R_e$ appears to reach a value close to zero within our error bars. It is obviously unphysical to interpret this simply in terms of Fermi volumes since this would imply an anomalously expanded $e$ pocket. The natural interpretation is that above $T \sim 50$~K the FS no longer contains an $e$ pocket. However, it is important to note from Fig.~3 that $\rho_{xy}(H)$ is visibly nonlinear at $T = 50$ and 60~K. Moreover, the MR retains its non-$H^2$ two-carrier form at all temperatures studied. This observation, more than any other, suggests that there are two types of carriers that persist to $T > 50$~K. The decrease in $R_e$ with increasing $T$ is then most readily understood by recalling that for a two-dimensional metal, $\rho_{xy}$ is a sensitive function of FS curvature.\cite{Ong91} In a situation where $\ell$ is anisotropic and the FS has both positive and negative curvature, $\rho_{xy}$ can vary significantly from its expected isotropic Drude value, and even change sign as a function of $T$, as seen, for example, in heavily overdoped LSCO.\cite{Narduzzo08} Our data thus suggest that the remnant FS associated with $R_e(T)$ has both electron and hole character (from a transport perspective), a feature that places strong constraints on its topology.

\begin{figure}
\includegraphics[width=8.0cm,keepaspectratio=true]{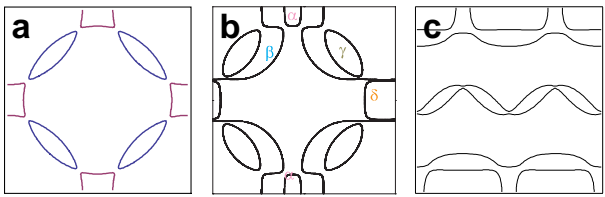}
\caption{(Color online) FS reconstruction of the CuO$_2$ plane due to (a) commensurate SDW or $d$DW order (Ref.~\onlinecite{Sachdev09}), (b) incommensurate SDW order (Ref.~\onlinecite{Harrison09}), and (c) antiphase stripe order (Ref.~\onlinecite{MillisNorman07}).} \label{Fig4}
\end{figure}

In light of these new findings, let us now turn to consider some of the possible scenarios, depicted schematically in Fig.~4, that may lead to the observation of small pockets and a negative $R_{\rm H}$ in Y-based cuprates. In both the commensurate SDW and $d$DW [Fig.~4(a)] as well as the incommensurate SDW [Fig.~4(b)] scenarios, the $h$ pockets are centered along the zone diagonals while the $e$ pockets appear at the zone boundaries. Close inspection of Fig.~4(a) reveals that the $e$ pockets in the commensurate case can have negative curvature, as implied by $R_e(T)$, though this is largely a result of the oversimplified circular FS used in such proposals. For a more realistic band structure for Y-based cuprates,\cite{Andersen95} such negative curvature vanishes upon ($\pi$, $\pi$) reconstruction, as indicated in Fig.~4(b).\cite{Harrison09} Moreover, it is not immediately clear why the mobility of the nodal carriers should undergo such a catastrophic change without there being a similarly adverse effect on the antinodal carriers.

Depending on the parametrization, antiphase stripe order can produce either a set of open, quasi-one-dimensional (1D) hole-like bands or $h$ pockets, in addition to $e$ pockets near the zone boundaries.\cite{MillisNorman07} Given the tendency of 1D bands to localize, the antiphase stripe scenario offers a natural explanation for the suppression of $\sigma_h$ below $T_h$, as well as the approach to zero of $R_h$ as $T \rightarrow 0$.\cite{Noda99} Such one dimensionality is also implicit in recent anisotropic Nernst measurements on Y123.\cite{Daou09} Moreover, the $e$ pocket can also have both positive and negative curvature, as shown in Fig.~4(c). Hence, on a qualitative level at least, the antiphase stripe scenario appears consistent with all our main experimental findings, though clearly, more detailed band-structure calculations on Y124 are needed to make a definitive statement on the feasibility of each scenario.

In summary, analysis of Hall and MR data on Y124 has revealed a number of important findings. First, the overall quality of the fits implies that a picture of coexisting $e$-like and $h$-like carriers is appropriate for Y-based cuprates [while we do not consider Y123 in this Rapid Communication, the evolution of the Hall and MR data are near identical to those found in Y124 (Ref.~\onlinecite{LeBoeuf07})]. Second, the quantitative agreement between the parameters obtained from our analysis and those derived from QO experiments confirms that the QO seen in UD Y-based cuprates arise from the same origin (Landau quantization) as those seen in the OD regime.\cite{Vignolle08} The dramatic fall in $\sigma_h$ at low $T$, however, reveals a marked change in the mobility of the holelike carriers, possibly due to their quasi-1D nature.\cite{Daou09} Finally, with the exception of the $d$DW case, all DW scenarios assume some sort of {\it field}-induced magnetic order, yet our fitting parameters are field independent. This implies that if the FS reconstruction is field induced, it must take place at fields $H < H_n$. In this regard, it would be interesting to learn whether the static (incommensurate) order inferred from recent low-$T$ neutron-scattering studies on UD Y123 (Ref.~\onlinecite{Haug09}) can be induced at arbitrarily low fields above $T_c$.

We thank A. Carrington and B.\ Fauqu\'{e} for helpful discussions. This work was supported by the EPSRC (U.K.), the French ANR DELICE, EuroMagNET (EU) under Contract No. RII3-CT-2004-506239, the Canadian Institute for Advanced Research and a Canadian Research Chair.


\end{document}